\documentclass[11pt,a4paper]{article}
\pdfoutput=1
\usepackage{jinstpub}
\toccontinuoustrue

\title{Commissioning of the MEG II tracker system}

\author[a,1]{M.~Chiappini,\note{Corresponding author.}}
\author[a]{A.~M.~Baldini,}
\author[c,d]{G.~Cavoto,}
\author[a,b]{F.~Cei,}
\author[c,d,1]{G.~Chiarello,}
\author[e]{A.~Corvaglia,}
\author[a,b]{M.~Francesconi,}
\author[a]{L.~Galli,}
\author[e]{F.~Grancagnolo,}
\author[a]{M.~Grassi,}
\author[h]{M.~Hildebrandt,}
\author[c,d]{M.~Meucci,}
\author[e]{A.~Miccoli,}
\author[a,b]{D.~Nicol\`o,}
\author[e,f]{M.~Panareo,}
\author[a,b,h]{A.~Papa,}
\author[a]{F.~Raffaelli,}
\author[c,g]{F.~Renga,}
\author[h,i]{P.~Schwendimann,}
\author[a]{G.~Signorelli,}
\author[e,f]{G.~F.~Tassielli,}
\author[c]{C.~Voena}

\affiliation[a]{INFN Sezione di Pisa, Largo B. Pontecorvo 3, 56127, Pisa, Italy}
\affiliation[b]{Dipartimento di Fisica dell'Universit\`a di Pisa, Largo B. Pontecorvo 3, 56127, Pisa, Italy}
\affiliation[c]{INFN Sezione di Roma, Piazzale A. Moro 2, 00185, Roma, Italy}
\affiliation[d]{Dipartimento di Fisica dell'Universit\`a ``Sapienza'' di Roma, Piazzale A. Moro 2, 00185, Roma, Italy}
\affiliation[e]{INFN Sezione di Lecce, Via per Arnesano, 73100, Lecce, Italy}
\affiliation[f]{Dipartimento di Matematica e Fisica dell'Universit\`a del Salento, Via per Arnesano, 73100, Lecce, Italy}
\affiliation[g]{Laboratori Nazionali di Frascati, Via Enrico Fermi 40, 00044, Frascati, Italy}
\affiliation[h]{Paul Scherrer Institut (PSI), Forschungsstrasse 111, 5232, Villigen, Switzerland}
\affiliation[i]{ETH Z\"{u}rich, R\"{a}mistrasse 101, 8092, Z\"{u}rich, Switzerland}

\emailAdd{marco.chiappini@pi.infn.it}

\abstract{The MEG experiment at the Paul Scherrer Institut (PSI) represents the state of the art in the search for the charged Lepton Flavour Violating (cLFV) $\mu^+ \rightarrow e^+ \gamma$ decay. With the phase 1, MEG set the new world best upper limit on the $\mbox{BR}(\mu^+ \rightarrow e^+ \gamma) < 4.2 \times 10^{-13}$ (90\% C.L.). With the phase 2, MEG II, the experiment aims at reaching a sensitivity enhancement of about one order of magnitude compared to the previous MEG result. The new Cylindrical Drift CHamber (CDCH) is a key detector for MEG II. CDCH is a low-mass single volume detector with high granularity: 9 layers of 192 drift cells, few mm wide, defined by $\sim 12000$ wires in a stereo configuration for longitudinal hit localization. The filling gas mixture is Helium:Isobutane (90:10). The total radiation length is $1.5 \times 10^{-3}$ $\mbox{X}_0$, thus minimizing the Multiple Coulomb Scattering (MCS) contribution and allowing for a single-hit resolution $< 120$ $\mu$m and an angular and momentum resolutions of 6 mrad and 90 keV/c respectively. This article presents the CDCH commissioning activities at PSI after the wiring phase at INFN Lecce and the assembly phase at INFN Pisa. The endcaps preparation, HV tests and conditioning of the chamber are described, aiming at reaching the final stable working point. The integration into the MEG II experimental apparatus is described, in view of the first data taking with cosmic rays and $\mu^+$ beam during the 2018 and 2019 engineering runs. The first gas gain results are also shown. A full engineering run with all the upgraded detectors and the complete DAQ electronics is expected to start in 2020, followed by three years of physics data taking.}


\begin{document}
\maketitle
\flushbottom

\section{Introduction}

The MEG experiment with its first phase of operation at the Paul Scherrer Institut (PSI) set the most stringent constraint on the charged Lepton Flavour Violating (cLFV) $\mu^+ \rightarrow e^+ \gamma$ decay. The analysis of the 2009--2013 full data set resulted in the new best upper limit on the $\mbox{BR}(\mu^+ \rightarrow e^+ \gamma) < 4.2 \times 10^{-13}$ at 90\% C.L. (ref.~\cite{finalresult}), imposing one of the tightest bounds on models predicting cLFV enhancements through new physics beyond the Standard Model (refs.~\cite{nuovocimento,cei}). The MEG experiment has reached its ultimate level of sensitivity, limited by the resolutions on the measurement of the kinematic variables of the two decay products (ref.~\cite{upgrade}). Therefore an upgrade of MEG, i.e. MEG II, was designed (ref.~\cite{meg2design}) and is presently in the commissioning phase at PSI. MEG II aims at reaching a sensitivity level of the order of $6 \times 10^{-14}$ in three years of data taking, by improving both, the detector resolutions and the muon stopping rate by a factor of two.

\section{The new MEG II Cylindrical Drift CHamber (CDCH)}

\subsection{Design}

\begin{figure}[h]
\centering
\includegraphics[width=.95\linewidth]{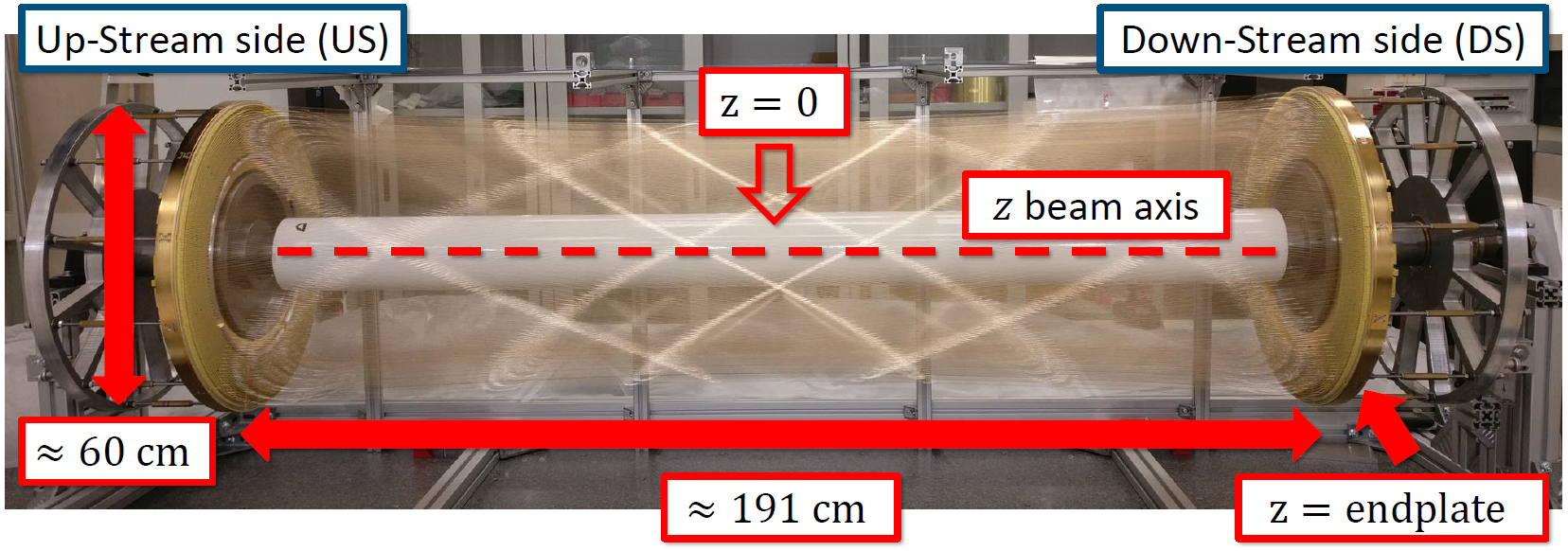}
\caption{The fully wired MEG II CDCH.}
\label{fig:cdchside}
\end{figure}

The new MEG II positron tracker is a single volume drift chamber with a cylindrical symmetry along the $\mu^+$ beam axis. The length is $\sim 191$ cm and the radial width ranges from $\sim 17$ to $\sim 29$ cm (figure~\ref{fig:cdchside}). The full azimuthal coverage around the $\mu^+$ stopping target is guaranteed. This improves the geometric acceptance for signal $\mbox{e}^+$ and allows to use new tracking procedures capable to exploit a factor of four more hits than MEG for a larger tracking efficiency ($\sim 70$\%). The high granularity is ensured by 9 layers of 192 drift cells each, few mm wide. Each layer consists of 2 criss-crossing field wire planes enclosing a sense wire plane. The wires are not parallel to CDCH axis, but form an angle varying from 6$^{\circ}$ in the innermost layer to 8.5$^{\circ}$ in the outermost one. The stereo angle has an alternating sign, depending on layer, allowing to reconstruct the longitudinal hit coordinate. The stereo configuration gives CDCH the shape of a rotation hyperboloid. The single drift cell is quasi-square with a 20 $\mu$m Au-plated W sense wire surrounded by 40/50 $\mu$m Ag-plated Al field wires, with 5:1 field-to-sense wires ratio and a total number of 11904 wires.

The sensitive volume is filled with a low-mass He:i$\mbox{C}_4\mbox{H}_{10}$ (90:10) gas mixture (ref.~\cite{gassystem}), which is a good compromise between high transparency (radiation length $\sim 1.5 \times 10^{-3}$ $\mbox{X}_0$) and single-hit resolution ($< 120$~$\mu$m, measured on prototypes, ref.~\cite{articoloSHR}). Full MC studies show angular and momentum resolutions in agreement with the MEG II experimental requirements: 6 mrad and 90 keV/c respectively.

\subsection{Construction}

\begin{figure}[h]
\centering
\includegraphics[width=.9\linewidth]{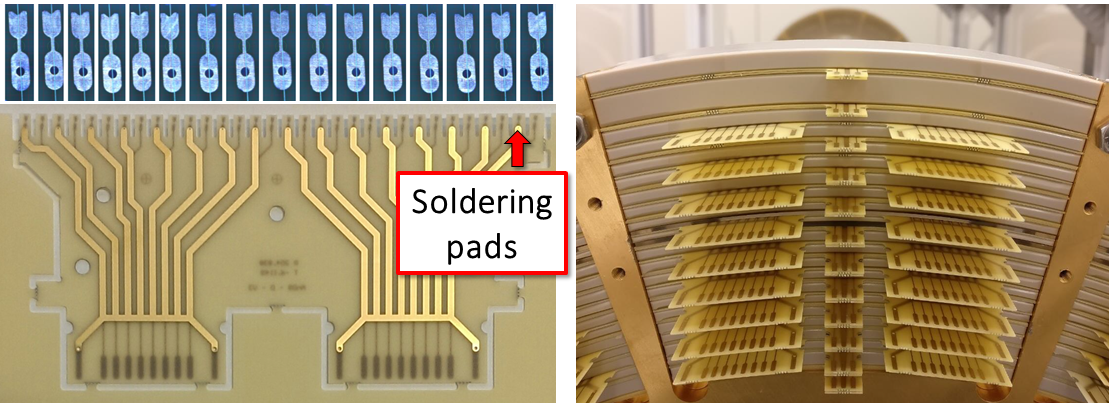}
\caption{Left: one of the PCBs where wires are soldered. The soldering pads are highlighted. Right: wire-PCBs stack with PEEK spacers between the spokes of the CDCH endplate.}
\label{fig:pcb}
\end{figure}

The CDCH design and construction were very challenging (refs.~\cite{elba1,elba2,vci}). Indeed, given the high wire density (12 wires/$\mbox{cm}^2$), the classical technique with wires anchored to endplates with feedthroughs is hard to implement. CDCH is the first drift chamber ever designed and built in a modular way. Wires were not strung directly on the final chamber, but they were soldered at both ends on the pads of two PCBs, which were then mounted on the chamber (figure~\ref{fig:pcb}, left). The wiring procedure was performed at INFN Lecce, exploiting an automatic robot which fixed the wires on PCBs with a contact-less laser soldering. CDCH was then assembled at INFN Pisa by radially overlapping the wire-PCBs in the twelve 30$^{\circ}$ sectors of the helm-shaped endplates, between the spokes which act as housing for the PCBs. Each wire-PCB was placed at the proper radius through PEEK spacers whose thickness was adjusted to have the correct radial dimension of the drift cells (figure~\ref{fig:pcb}, right).

\begin{figure}[h]
\centering
\includegraphics[width=.95\linewidth]{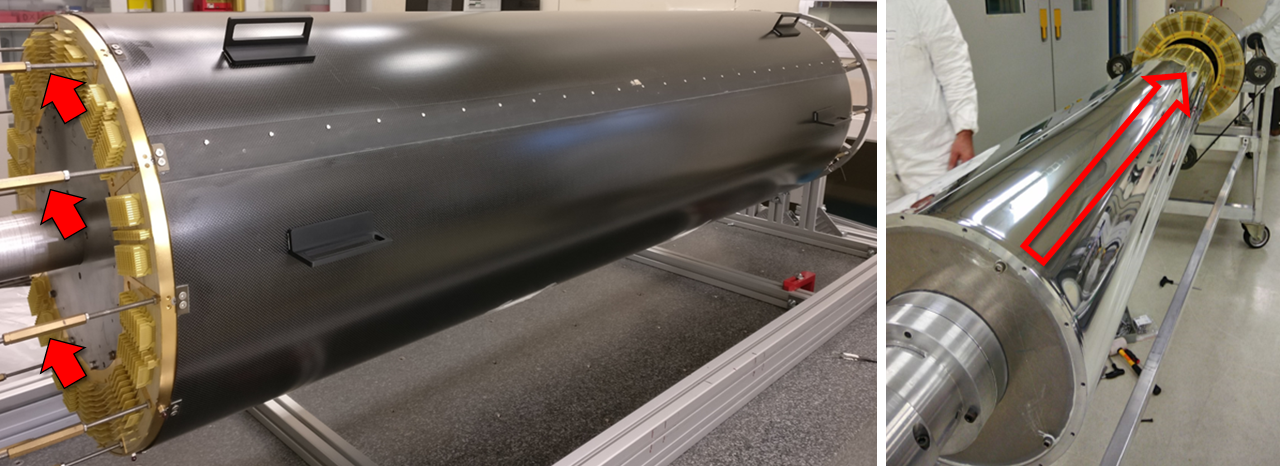}
\caption{Left: the CF support structure with some turnbuckles highlighted. Right: insertion phase of the 20 $\mu$m Al Mylar foil.}
\label{fig:sealing}
\end{figure}

At the outermost radius, a 2 mm-thick Carbon Fiber (CF) support structure encloses the sensitive volume and keeps the endplates at the correct distance, ensuring the proper mechanical tension of the wires (figure~\ref{fig:sealing}, left). Twelve turnbuckles per side were connected to each individual spoke, allowing a fine tuning of the endplates distance, planarity and parallelism at a level better than 100 $\mu$m. The CDCH geometry was continuously monitored during the assembly with a coordinate measuring machine. At the innermost radius, a 20 $\mu$m one-side-Al Mylar foil separates the CDCH gas volume from the He-filled target region (figure~\ref{fig:sealing}, right). The gas mixture tightness is achieved by using the ThreeBond 1530\footnote{https://www.threebond.co.jp/en/product/detail/tb1530.html. It is a single-component silicone with strong adhesion. It ensures a very good tightness, even for He.} glue and Stycast 2850\footnote{https://www.henkel-adhesives.com/it/en/product/potting-compounds/loctite\_stycast\_2850ft.html. It is a two-component thermally conductive epoxy encapsulant.} resin.

Aluminum wire breaking problems arose during the CDCH assembly and commissioning, despite the fact that all the operations were performed inside cleanrooms with a strict monitoring of the environmental conditions. The problem was deeply investigated performing optical inspections with microscopes, chromatography, practical tests and SEM/EDS analyses. We developed a safe procedure to extract the broken wire pieces from the chamber. By simulating the drift cells electric field with Garfield\footnote{https://garfieldpp.web.cern.ch/garfieldpp/.} and ANSYS\footnote{https://www.ansys.com/.}, the effect of a missing cathode wire on the $e^+$ reconstruction was found to be totally negligible. Chemical and mechanical analyses showed that the origin of the breaking phenomenon is the chemical corrosion of the Al core in presence of water condensation on wires from ambient humidity. Keeping the wires volume in an absolutely dry atmosphere with a continuous flow of inert gas (Nitrogen or Helium) proved to be effective to stop the development of corrosion. Once the assembly was completed, CDCH was transported to PSI for the commissioning phase.

More details about the CDCH design, construction and Al wires corrosion can be found in ref.~\cite{phd}.

\section{CDCH commissioning at PSI}

\subsection{Endcaps preparation}

\begin{figure}[h]
\centering
\includegraphics[width=.84\linewidth]{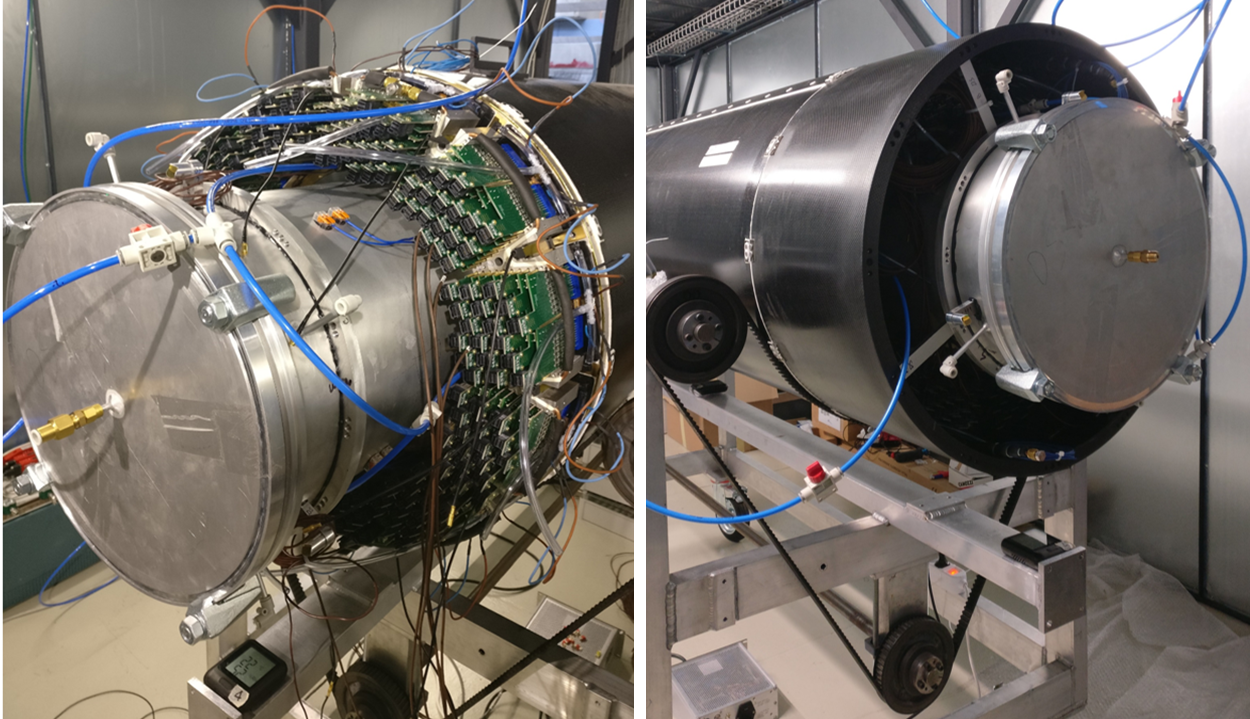}
\caption{Left: 216 FE boards connected to wire-PCBs tails (figure~\ref{fig:pcb}, right) and kept in position by the Al holders. One Al inner extension cylinder is visible. Right: one endcap region enclosed by the CF outer extension cylinder.}
\label{fig:endcap}
\end{figure}

After the CDCH sealing, the next step was the preparation of the endcap services. Two Al inner extension cylinders were connected to the endplates to couple the inner CDCH volume to the MEG II beam line. Twelve Al holders per side with grooves at the correct radii were machined to keep in position the 216 Front-End (FE) boards per side (figure~\ref{fig:endcap}, left).

\begin{figure}[h]
\centering
\includegraphics[width=.84\linewidth]{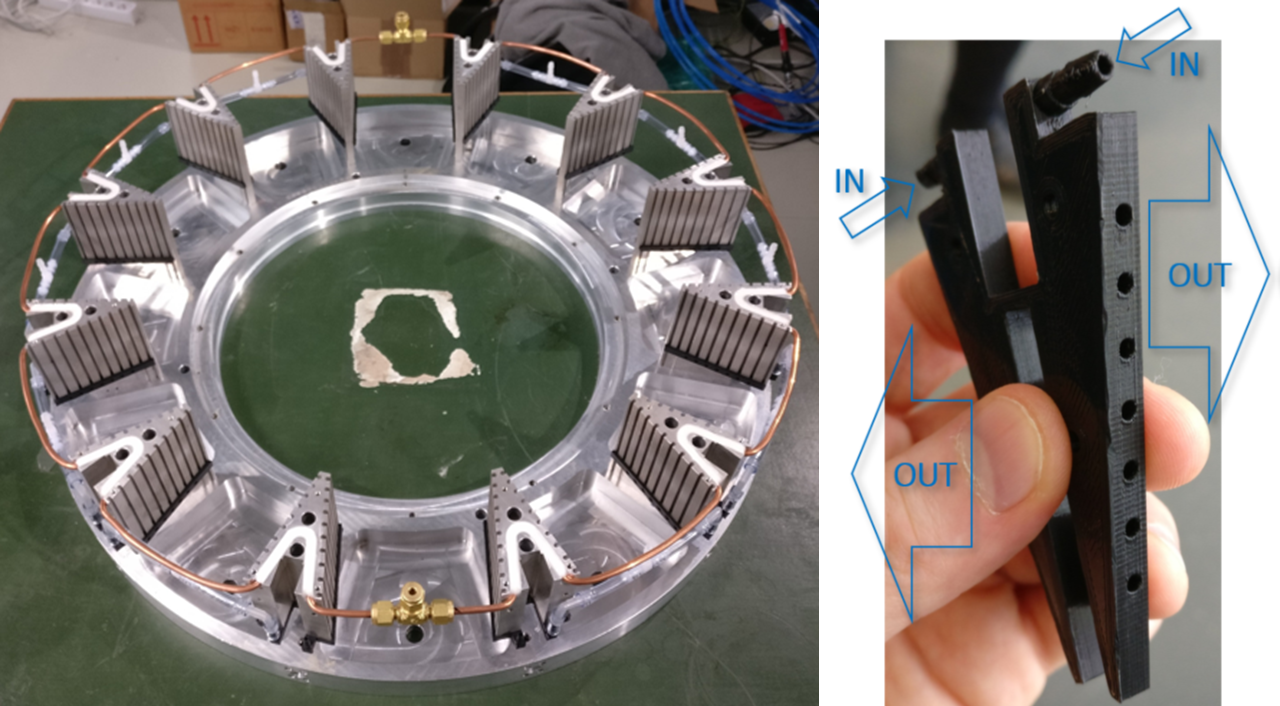}
\caption{Left: the 4 mm Cu pipes of the liquid cooling system directly embedded in the FE holders. Right: one 3D-printed piece used to flush dry air inside the endcaps.}
\label{fig:cooling}
\end{figure}

The FE boards can supply the HV to the wires from one side and read out the signal at both wire ends. In order to carry away the heat generated by the active electronics ($\sim 300$ W/endplate) a chiller system is used. The 4 mm Cu pipes of the liquid cooling system are directly embedded in the FE holders (figure~\ref{fig:cooling}, left). A system of twelve inlet tubes per side supplies dry air to 3D-printed pieces screwed on the back side of each holder (figure~\ref{fig:cooling}, right). The dry air flushing system is used to uniform the temperature inside the endcaps and avoid water condensation on the FE electronics. Temperature and humidity sensors were added to monitor the endcaps environment which is enclosed by CF outer extension cylinders (figure~\ref{fig:endcap}, right).

\subsection{Final working point and integration into the MEG II experimental apparatus}

The HV Working Point (WP) for CDCH was estimated through gas gain simulations with the He:Isobutane 90:10 mixture and typical atmospheric pressure values at PSI. Since we aim to be sensitive to the single ionization cluster, the WP was defined as the HV to get a gas gain $G = 5 \times 10^{5}$. A HV tuning by 10 V/layer was considered to compensate for the variation of the cell dimensions with radius. Furthermore, given the stereo wires geometry, the cell dimensions also vary along the chamber axis. The average HV WP as a function of the cell layers is summarized in table~\ref{tab:wp}.

\begin{table}[h]
\centering
\caption{HV working point as a function of the drift cell layers (L1 outermost, L9 innermost).}
\label{tab:wp}
\bigskip
\begin{tabular}{|c|c|c|c|c|c|c|c|c|}
\hline
L1 & L2 & L3 & L4 & L5 & L6 & L7 & L8 & L9 \\
\hline
1480 V & 1470 V & 1460 V & 1450 V & 1440 V & 1430 V & 1420 V & 1410 V & 1400 V \\
\hline
\end{tabular}
\end{table}

\begin{figure}[h]
\centering
\includegraphics[width=.99\linewidth]{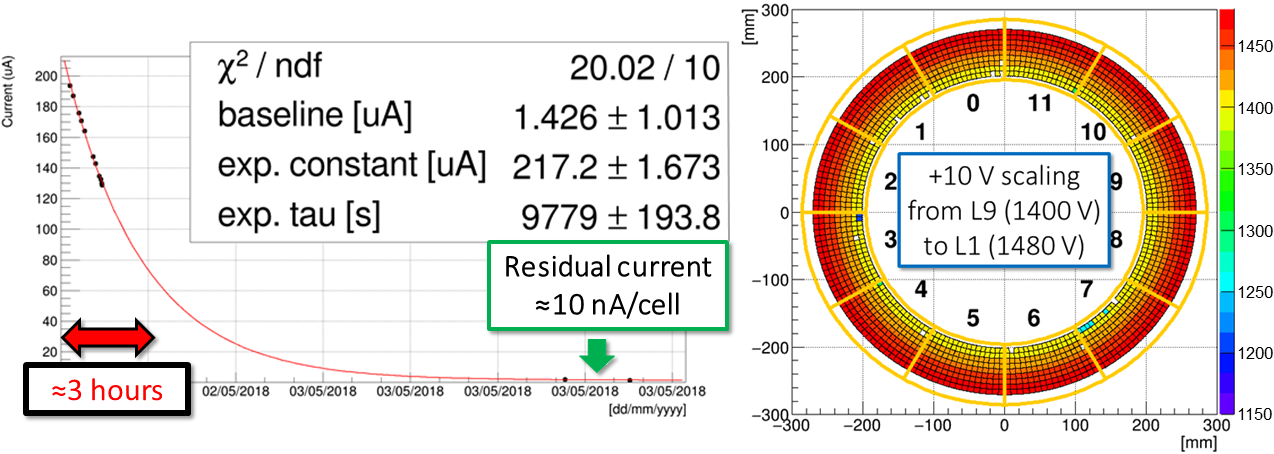}
\caption{Left: example of HV conditioning of the chamber at the first power up. Right: final HV map at the working point. The color scale ranges from 1150 V to 1480 V.}
\label{fig:conditioningmap}
\end{figure}

Figure~\ref{fig:conditioningmap} (left) shows an example of HV conditioning of the chamber at the first power up (here at 700 V). The residual currents drawn by the HV channels to correctly polarize the dielectric materials of the endplates reached a value of $\sim 10$ nA/cell, starting with a value more than a factor of 300 higher. The characteristic time was about three hours.

The CDCH working length was experimentally determined through systematic HV tests at different lengths/wires elongations, adjusted through geometry survey campaigns with a laser tracker ($\sim 20$ $\mu$m accuracy). The final length was set to +5.6 mm of wires elongation with respect to the zero tension position, corresponding to $\sim 70\%$ of the elastic limit. This guarantees an electrostatic stability safety margin of $\sim 100$ V above the WP. The map showing the HV value reached by each drift cell is reported in figure~\ref{fig:conditioningmap} (right). The cells which did not reach the WP are considered inefficient. The measured cell inefficiency was $1.3\%$, negligible in $e^+$ reconstruction.

\begin{figure}[h]
\centering
\includegraphics[width=.99\linewidth]{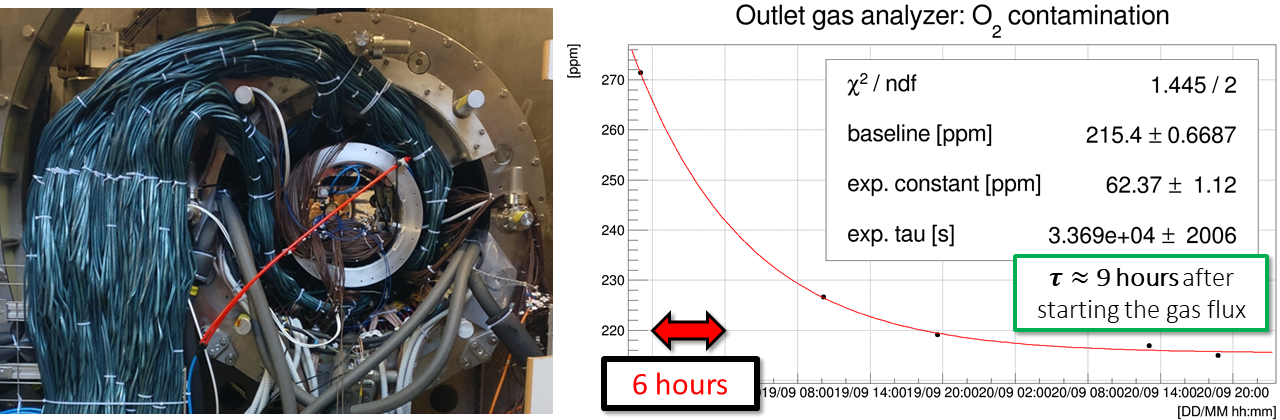}
\caption{Left: CDCH fully integrated into the MEG II experimental apparatus. The signal cables are visible, together with the gas/cooling system pipes. Right: O$_2$ contamination vs. time at the gas outlet.}
\label{fig:integrationoxygen}
\end{figure}

CDCH was finally integrated into the MEG II experimental apparatus for the first time in 2018, as shown in figure~\ref{fig:integrationoxygen} (left). Complete signal/HV cabling and gas inlet/outlet connections to the final MEG II gas system were performed. A gas analyzer at the chamber outlet monitors the contaminants. An example for the Oxygen content\footnote{Electronegative impurities, like O$_2$, can affect the $e^-$ avalanche development inside the drift cell, causing gas gain fluctuations. This is the so-called electron attachment.} as a function of time since the starting of the gas mixture flux is shown in figure~\ref{fig:integrationoxygen} (right). The cooling pipes were also routed and the cooling system was successfully tested. After the 2018 and 2019 engineering runs the CDCH geometry was measured and the CDCH mechanics proved to be stable and adequate to sustain a MEG II run.

\subsection{First data taking}

After completing the beam line and starting the gas mixture flux, CDCH was ready for data taking. Only 192 DAQ channels were available for 2018 and 2019 runs. This corresponds to six layers in one sector (16 drift cells per layer) with the double-side read out. Due to the limited number of DAQ channels a particle tracking test was not possible. Nevertheless, we studied the noise level in the experimental environment and performed several HV scans around the WP with Cosmic Rays (CR) and with the $\mu^+$ beam at different intensities: $\sim 10^{7}$ stopped $\mu^+$/s (low rate), $3 \times 10^{7}$ stopped $\mu^+$/s (rate during the phase 1 of MEG), $7 \times 10^{7}$ stopped $\mu^+$/s (rate planned for MEG II). CR data allowed the first experimental gas gain studies in a clean environment. Michel $e^+$ data allowed to test the chamber response in a high rate environment and the combined data taking with the other MEG II sub-detectors (ref.~\cite{meg2design}).

\begin{figure}[h]
\centering
\includegraphics[width=.95\linewidth]{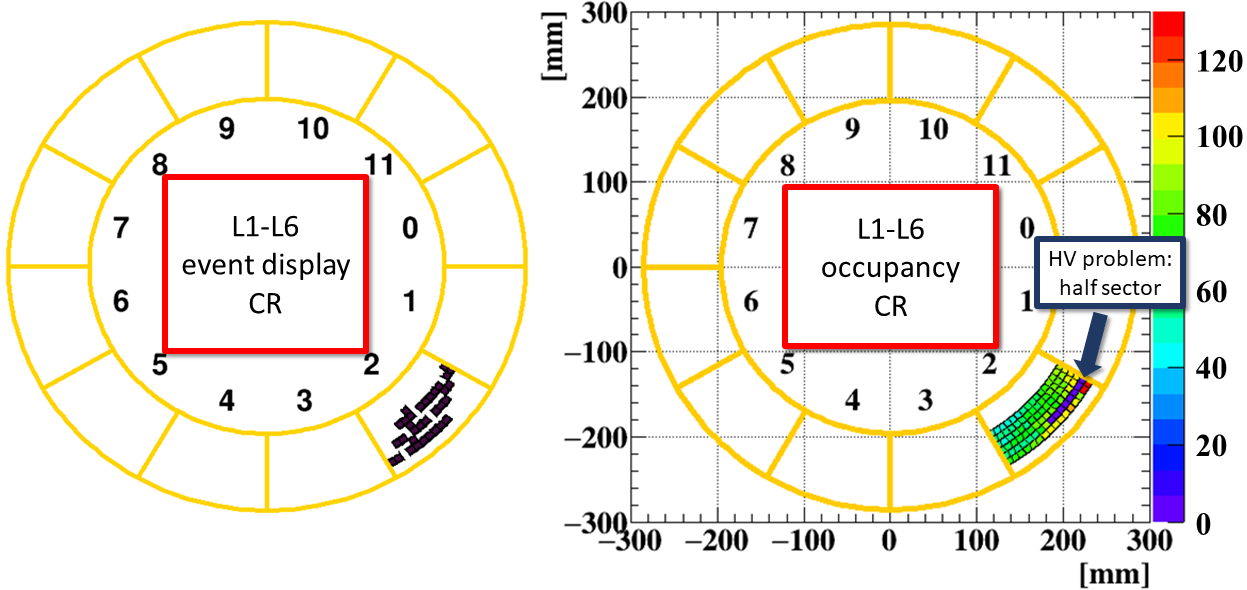}
\caption{Left: CR event display at the HV WP. Right: occupancy plot integrating 25000 events.}
\label{fig:crevent}
\end{figure}

\begin{figure}[h]
\centering
\includegraphics[width=.95\linewidth]{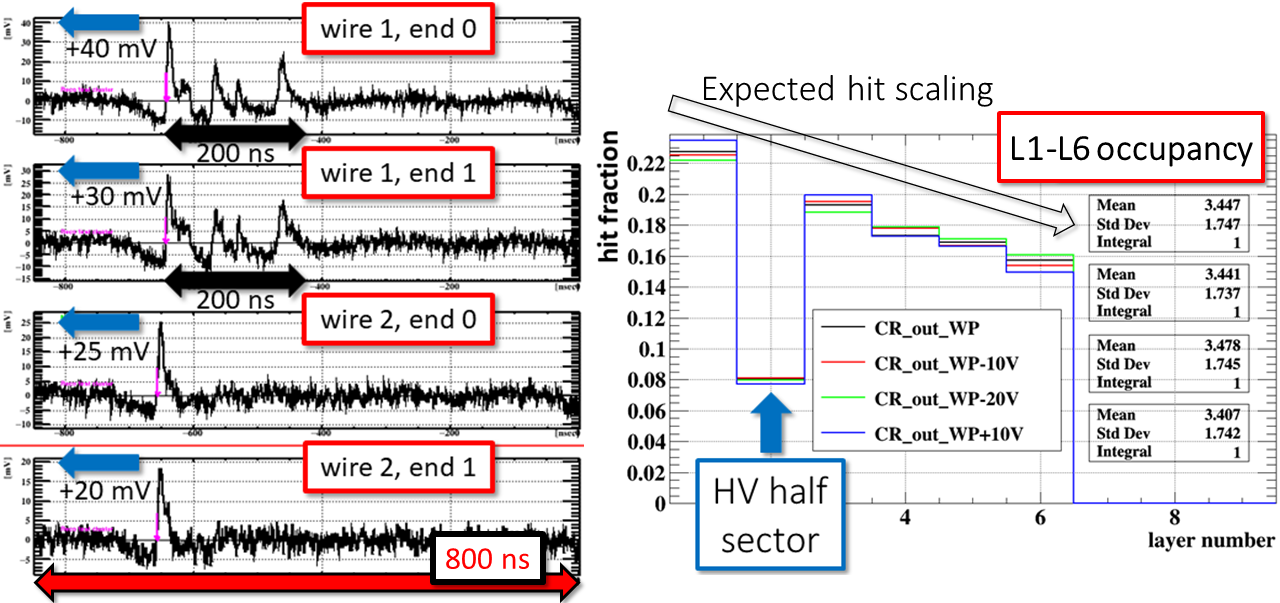}
\caption{Left: typical WFs as measured at both ends of two adjacent cells. Right: L1-L6 hit occupancy.}
\label{fig:wfoccupancy}
\end{figure}

\begin{figure}[h]
\centering
\includegraphics[width=.95\linewidth]{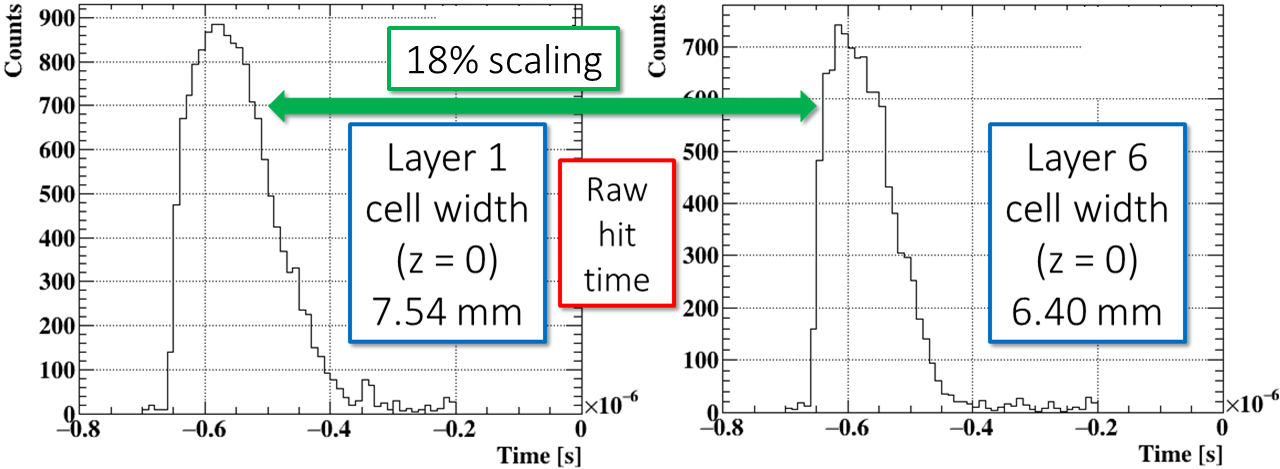}
\caption{Hit time distributions for L1 (left) and L6 (right). The $\sim 18\%$ width scaling is related to the different drift cell dimensions.}
\label{fig:hittime}
\end{figure}

\begin{figure}[h]
\centering
\includegraphics[width=.9\linewidth]{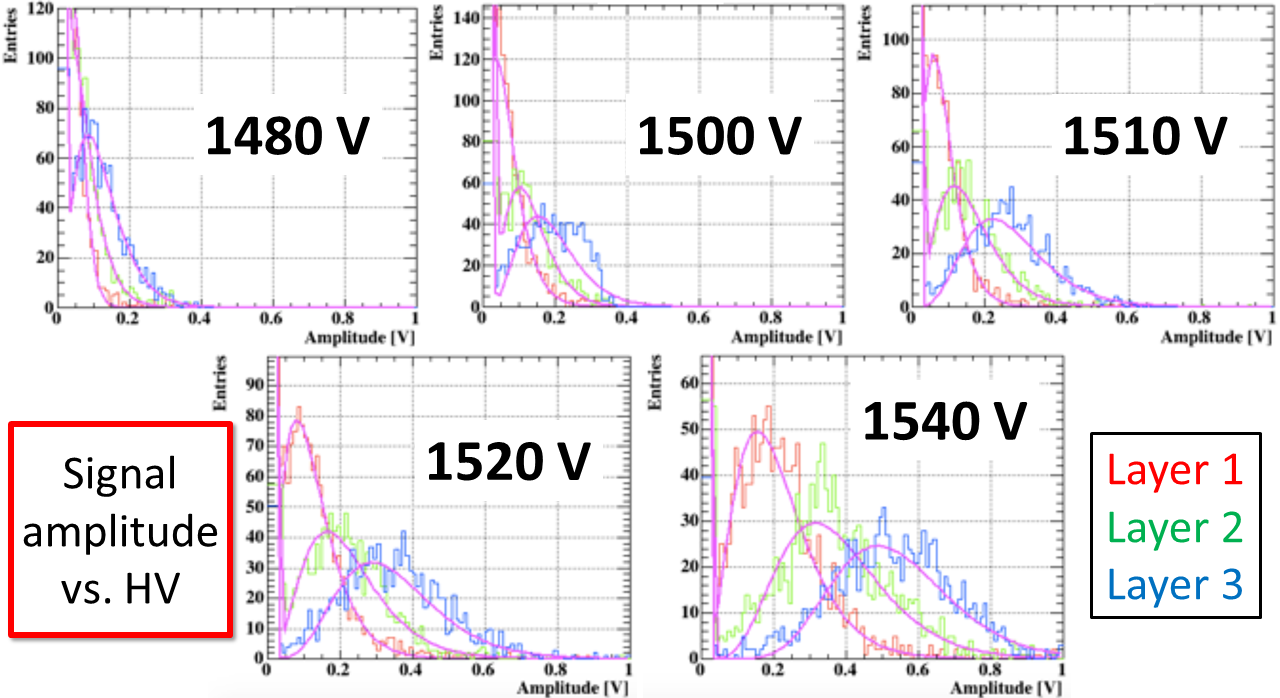}
\caption{Distribution of the signal WF amplitude vs. HV for L1, L2, L3 from CR data. These plots were used to extract the first gas gain estimate (figure~\ref{fig:gain} left).}
\label{fig:amplitude}
\end{figure}

Figure~\ref{fig:crevent} shows an example of a CR event display at the HV WP (left) from layer 1 (L1, outer) to layer 6 (L6, inner) and the corresponding occupancy plot integrating 25000 events. One drift cell has a hit if the WaveForm (WF) exceeds a predefined threshold: typically $\times 5$ the RMS of the noise baseline which was $\sim 2$ mV. Figure~\ref{fig:wfoccupancy} shows typical signal WFs as measured at both ends of two adjacent cells (left) and the occupancy as a function of the first six layers (right). The hit scaling as expected from MC simulations was correctly observed. Another interesting plot is reported in figure~\ref{fig:hittime} which shows the $\sim 18\%$ width scaling of the raw hit time distributions for L1 (left) and L6 (right). This is related to the different drift cell dimensions: 7.54 mm and 6.40 mm for L1 and L6 respectively at the CDCH center ($z = 0$, figure~\ref{fig:cdchside}). Figure~\ref{fig:amplitude} shows the distribution of the signal WF amplitude from CR data as a function of the same HV applied to L1, L2, L3.

\begin{figure}[h]
\centering
\includegraphics[width=.99\linewidth]{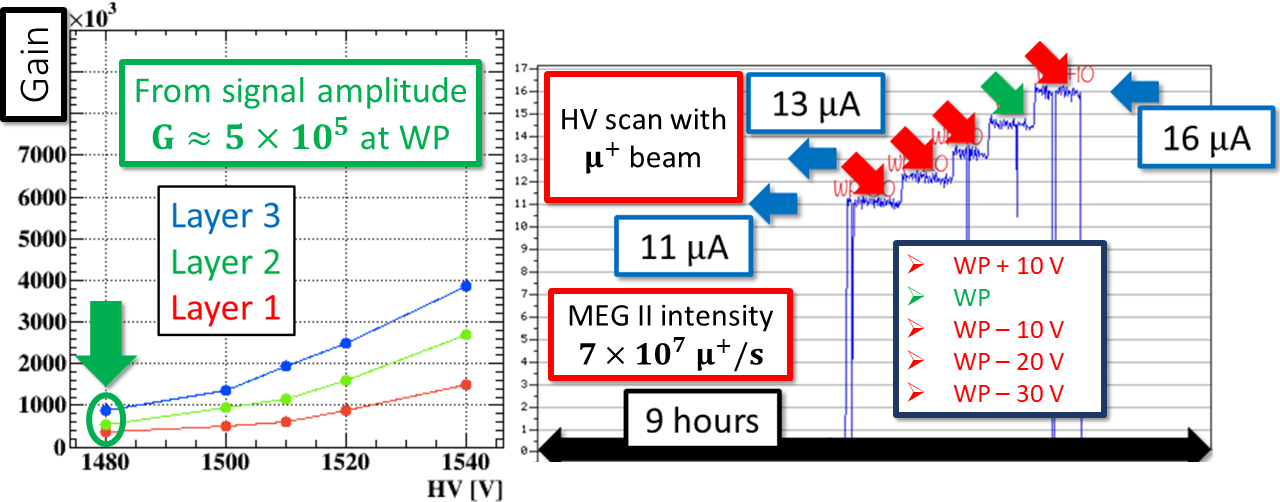}
\caption{Experimental gain curves as extracted from CR data (left) and from a HV scan with $\mu^+$ beam at the MEG II intensity (right).}
\label{fig:gain}
\end{figure}

Distributions were fitted with a gaussian pedestal + P\'{o}lya distribution\footnote{Typical shape from the avalanche statistics.} for signal. The mean amplitude and thus the separation from pedestal increase as the HV was set to higher values. The mean amplitude is higher for L3 than L1 at fixed HV, given the higher gain for inner layers (smaller cells). This is the reason for the HV scaling shown in table~\ref{tab:wp}. The mean amplitude was then converted into the effective gas gain $G$ by means of simulations of the ionization clusters and the response of the FE amplification stage. The first calibrated gain curves, as extracted from CR data, are reported in figure~\ref{fig:gain} (left), showing agreement with simulation at the HV WP.

As aforementioned, also $\mu^+$ beam data were collected. Figure~\ref{fig:gain} (right) shows the current drawn by a HV channel in [$\mu$A] as a function of the HV applied around the WP (from WP~-~30~V up to WP~+~10~V in step of 10~V) at a beam intensity of $7 \times 10^{7}$ $\mu^+$/s (nominal MEG II rate). The experimental gain curve showed a nearly exponential behaviour as expected from gas gain simulations.

During the 2018 and 2019 engineering runs we experienced anomalous high current levels in some sectors and layers, starting with the $\mu^+$ beam. The complete power down of the chamber was necessary to drive the big currents to zero. This behaviour needs to be carefully understood.

At present the commissioning phase at PSI is still ongoing. A full engineering run with all the upgraded detectors and the complete DAQ electronics is expected to start in 2020, followed by three years of physics data taking.

More details about the CDCH commissioning and first data taking can be found in ref.~\cite{phd}.

\end{document}